\documentclass[10pt]{article}
\usepackage[english]{babel}
\usepackage{amsfonts}
\usepackage[dvips]{graphicx}

\begin{document}

\title{\bf An expansion based on Reynolds number powers for the velocity field of elementary analytical flows}

\date{September 2006}

\author{\bf Gianluca Argentini \\
\normalsize gianluca.argentini@riellogroup.com \\
\textit{Research \& Development Department}\\
\textit{Riello Burners}, 37048 San Pietro di Legnago (Verona), Italy}

\maketitle

\begin{abstract}
The work describes a possible physical characterization for the definition of elementary fluid flow. As consequence, an analytical expansion based on Reynolds number powers for the velocity field is shown in the weakly turbulent case.
\end{abstract}

\section{Notion of elementary analytical flow}

Let us recall the notion of elementary analytical flow (see \cite{argentini}).\\
Let $L$ a length unit and $\bar{\Omega}$ a closed domain of $\mathbb{R}^n$, $n \geq 2$, formed by a small number (e.g. 2 or 3) of $L^2$ squares or $L^3$ cubes. In a 2D steady case, let $F$ a fluid entering in $\Omega$ from a single edge of a square and flowing out from another single edge. Let also $s$, $0 \leq s \leq 1$, a parameter which describes the position of a single fluid particle along the inflow edge, so that $s_0L$ is the initial position of the generic particle associated to the particular value $s_0$ of the parameter.\\
Let $s \mapsto R(s)$ a function $R:[0,1] \rightarrow [0,1]$ which maps a value of the parameter $s$ to the value identifying the position reached by the particle on the outflow edge, so that this position is identified by the value $R(s)L$. If ${x,y}$ is a cartesian coordinates system in the plane of $\Omega$, assume that each streamline, or each particle path in lagrangian view, is described by a parametric curve $\tau \mapsto \Phi(\tau)=\left(x(\tau),y(\tau)\right)$, with $0 \leq \tau \leq 1$. The parameter $\tau$ could not be, in general, the time variable of the flow. Let this parametric representation be determined by a set $A$ of analytical conditions regarding $\Phi(\tau)$ and $\dot{\Phi}(\tau)$, that is the passage of the particle in some suitable points of $\Omega$ and the velocity field direction in some other (or the same) points.\\
An {\it elementary analytical flow} is then a particular list $\mathbb{F}=\left\{\Omega,R,A\right\}$.\\

\noindent In this work let $\Omega$ the domain formed by three $L^2$ squares, the first two along the $x$-axis of a cartesian coordinates system from $0$ to $2L$, and the third above the second, from $L$ to $2L$ $y$-coordinates (\cite{argentini}).\\
\noindent Now the set $A$ af analytical conditions is so defined:\\
\indent $P_1$. at the inflow edge, for $\tau=0$ and position parameter $s$, let $\Phi(0)=(0,sL)$;\\
\indent $P_2$. at the outflow edge, for $\tau=1$ and position parameter $R=R(s)$, let $\Phi(1)=(L(1+R),2L)$;\\
\indent $P_3$. for $\tau=\frac{1}{2}$ a particle path intersects the diagonal line of the second square, that is the line of cartesian equation $y=-x+2L$; if a parameter $p$, $0 \leq p \leq 1$, describes the positions along this line, the condition is $\Phi(\frac{1}{2})=(L(2-p),pL)$;\\
\indent $D_1$. at the inflow edge the velocity is parallel to $x$-axis, so that for every $s$ $\dot{y}(0)=0$;\\
\indent $D_2$. at the outflow edge the velocity is parallel to $y$-axis, so that for every $s$ $\dot{x}(1)=0$.\\
\noindent Assuming that the components of $\Phi(\tau)$ are cubic polinomials on variable $\tau$, that is the more simple analytical expressions satisfying the set of conditions $A$, by usual computations follows (\cite{argentini})

\begin{eqnarray}\label{components}
	x(\tau)=L(10-6R-8p)\tau^3+L(-21+11R+16p)\tau^2+L(12-4R-8p)\tau\\
	y(\tau)=L(4+6s-8p)\tau^3+L(-2-7s+8p)\tau^2+Ls
\end{eqnarray}

\noindent Also, let $p=s$, so that particles paths have their initial shape until the line $y=-x+2L$. In this case the curve $\Phi$ has the form

\begin{eqnarray}\label{curveps}
	x(\tau)=L(-6R-8s+10)\tau^3+L(11R+16s-21)\tau^2+L(-4R-8s+12)\\
	y(\tau)=L(2-s)(2\tau^3-\tau^2)+Ls
\end{eqnarray}

\noindent Assume now the following definition for $R(s)$, which gives a possible mathematical modelization of the physical phenomenon of weak turbulence (see \cite{argentini} and, e.g., \cite{roux})

\begin{equation}\label{weakTurbulent}
	R(s) = \frac{1}{2}\left[1+sin\left(1-\frac{2}{c_{\mathbb{F}}}\hspace{0.1cm}\mathbb{R}e\hspace{0.1cm}s\right)\right]
\end{equation}

\noindent where $\mathbb{R}e$ is the Reynolds number and $c_{\mathbb{F}}$ is a constant associated to the flow. The previous form of $R(s)$ is a consequence of the following Taylor expansion

\begin{equation}\label{ReLimit}
	\lim_{\mathbb{R}e \rightarrow c_{\mathbb{F}}}R(s)=\frac{1}{2}\left[1+sin\left(1-2s\right)\right] \approx 1-s
\end{equation}

\noindent which is the correct shape of $R(s)$ in the laminar case (\cite{argentini}).

\section{A possible expression for $c_{\mathbb{F}}$}

The relation (\ref{ReLimit}) is the viscous limit (the inverse of the {\it inviscid limit}, see (\cite{constantin})), which is valid for small values of Reynolds number. On the contrary, when $\mathbb{R}e$ is large, the argument of the sine function in (\ref{weakTurbulent}) should assume great values for the {\it frequency} $\frac{\mathbb{R}e}{c_{\mathbb{F}}}$, so that the flow becomes turbulent because, for a fixed particle, $R(s)$ can be a value very different compared to laminar value $1-s$.\\
If we would fixed the constant $c_{\mathbb{F}}$ indipendently from the flow considered, that is from the analytical representation $\Phi(\tau)$, the most simple choice which satisfy the two previous asymptotic constraints is

\begin{equation}\label{cF}
	c_{\mathbb{F}}=\frac{1}{\mathbb{R}e}
\end{equation}

\noindent Substituting (\ref{cF}) into (\ref{weakTurbulent}), the viscous limit (\ref{ReLimit}) is expressed by the relation

\begin{equation}\label{RecF}
	\lim_{\mathbb{R}e \rightarrow c_{\mathbb{F}}} \mathbb{R}e^2 = 1
\end{equation}

\noindent that is $c_{\mathbb{F}}=\mathbb{R}e=1$, which can be viewed as a physical definition of viscous limit for elementary flows.\\
A more general choice could be

\begin{equation}\label{cFgeneral}
	c_{\mathbb{F}}=\frac{1}{\sum_{i=0}^NA_i\mathbb{R}e^i}
\end{equation}

\noindent but in this case we should estimate the coefficients $A_i$, and the value of $\mathbb{R}e$ in the case of viscous limit could be not unique.

\section{Choice of $c_{\mathbb{F}}$ and time parameterization}

In this section we show a relation between $c_{\mathbb{F}}$ and time parameterization for an elementary flow.\\
Assume that the parameter $\tau$ of the paths representation $\Phi(\tau)$ could depend, at least as first approximation, linearly on time variable $t$ by a multiplicative regular function $f=f(s)$, where $s$ is the parameter identifying a single particle path from inflow position:

\begin{equation}\label{tauParameter}
	\tau=f(s)t
\end{equation}

\noindent In this way, the flow velocity field $(u_1(s,t),u_2(s,t))$ can change from one path to another. Note that, for a fixed value of $s$, this formula has meaning only for $0 \leq t \leq f(s)^{-1}$. For every fixed $s$ follows

\begin{eqnarray}\label{timeDerivation}
	u_1=\frac{dx}{dt}=\frac{dx}{d\tau}\frac{d\tau}{dt}=f(s)\dot{x}\\
	u_2=\frac{dy}{dt}=\frac{dy}{d\tau}\frac{d\tau}{dt}=f(s)\dot{y} \nonumber
\end{eqnarray}

\noindent Now, let $t=0$. Then, being in the considered example $\dot{y}(0)=0$ $\forall s$, the initial velocity field at inflow edge, usually a known boundary condition, is $(u_1(s,0),u_2(s,0))=(u_1(s,0),0)$. From (\ref{curveps}) follows that $\dot{x}(0)= 4L(3-R-2s)$, and at least for those inflow points for which $R(s)\neq3-2s$ the analytical expression of $f(s)$ is

\begin{equation}\label{fs}
	f(s)=\frac{u_1(s,0)}{\dot{x}(0)}
\end{equation}

\noindent The analytical relation between velocity field and particles paths is then

\begin{equation}\label{velField}
	(u_1(s,t),u_2(s,t))=\frac{u_1(s,0)}{\dot{x}(0)}\left(\dot{x}(\tau),\dot{y}(\tau)\right)_{|\tau=\frac{u_1(s,0)}{\dot{x}(0)}t}
\end{equation}

\section{An expansion for velocity field}

\noindent Using (\ref{cF}), suppose now that Reynolds number $\mathbb{R}e$ is small enough to have $1-2\mathbb{R}e^2s \approx 0$, so that at first order $R(s) \approx 1-\mathbb{R}e^2s$. Then $\dot{x}(0)=4L[2+(\mathbb{R}e^2-2)s]$. Also, assuming regular properties for $u_1(s,0)$, from Taylor expansion centered e.g. on $s=0$ let $u_1(s,0) \approx u_{10}+\dot{u}_{10}s+\frac{1}{2}\ddot{u}_{10}s^2$, where $u_{10}=u_1(0,0)$, $\dot{u}_{10}=\partial_s u_1(0,0)$ and $\ddot{u}_{10}=\partial^2_{ss} u_1(0,0)$. Using a Taylor expansion for $f(s)$, it can be shown that

\begin{eqnarray}\label{fsExpansion}
f(s) \approx \frac{1}{8L}\left[u_{10}\right]+\frac{1}{8L}\left[\left(u_{10}+\dot{u}_{10}-\frac{\mathbb{R}e^2}{2}u_{10}\right)s\right]+ \\
\frac{1}{8L}\left[\left(u_{10}+\dot{u}_{10}+\frac{\ddot{u}_{10}}{2}-\mathbb{R}e^2u_{10}-\frac{\mathbb{R}e^2}{2}\dot{u}_{10}+\frac{\mathbb{R}e^4}{4}u_{10}\right)s^2\right] \nonumber
\end{eqnarray}

\noindent This formula shows the role of Reynolds number and inflow boundary condition about the time parameterization (\ref{tauParameter}) of the particles paths and about the expression of velocity field ((\ref{fs})+(\ref{velField})). For high values of $\mathbb{R}e$, the expansion of $R(s)$ should consider higher order terms, so that the approximated expression for the velocity field could present higher power of Reynolds number (for a method of expansion based on $\mathbb{R}e$ powers see (\cite{lundgren}). Usual computations give the following series expansion for $f(s)$, in the case of a parabolic inflow velocity $u_1(s,0)=v_0+v_1s+v_2s^2$:

\begin{eqnarray}\label{seriesExpansion}
f(s)= \frac{v_0}{8L}+\frac{1}{8L}\left[\frac{(2-\mathbb{R}e^2)}{2}v_0+v_1\right]s+ \\
+\frac{1}{8L}\sum_{m=2}^{+\infty}\left[\frac{(2-\mathbb{R}e^2)^mv_0}{2^m}+\frac{(2-\mathbb{R}e^2)^{m-1}v_1}{2^{m-1}}+\frac{(2-\mathbb{R}e^2)^{m-2}v_2}{2^{m-2}}\right]s^m \nonumber
\end{eqnarray}

\noindent Using the two relations (\ref{timeDerivation}), previous formula gives an expansion of velocity field in terms of Reynolds number powers. In the attached figures we present some profiles for the scalar speed $\sqrt{u_1^2+u_2^2}$ in the case of an initial Poiseuille flow $u_1(s,0)=5(s-s^2)$.

\begin{figure}
	\begin{center}
	\includegraphics[width=13cm]{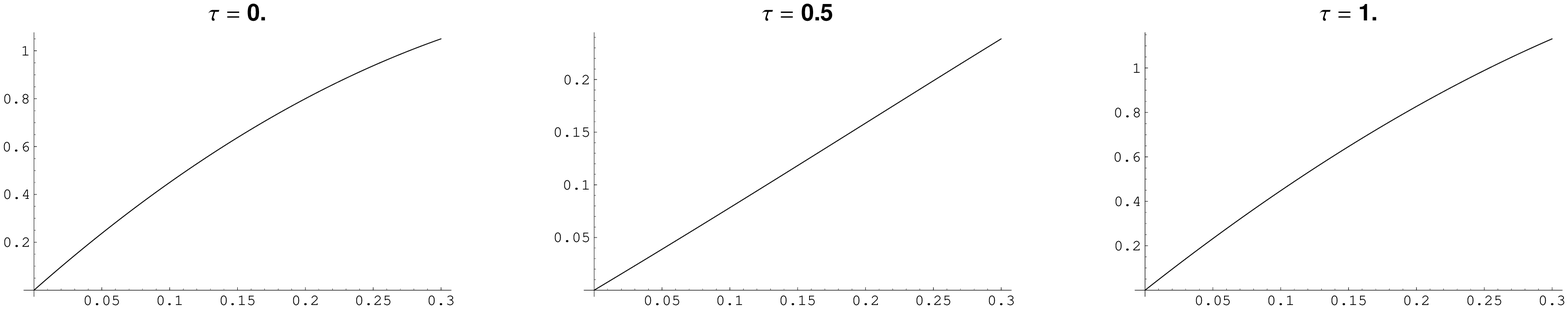}
	\caption{\small{\it Profiles of scalar speed for three value of $\tau$, $0 \leq s \leq 0.3$ and $\mathbb{R}e=1$.}} 
	\end{center}
\end{figure}

\begin{figure}
	\begin{center}
	\includegraphics[width=13cm]{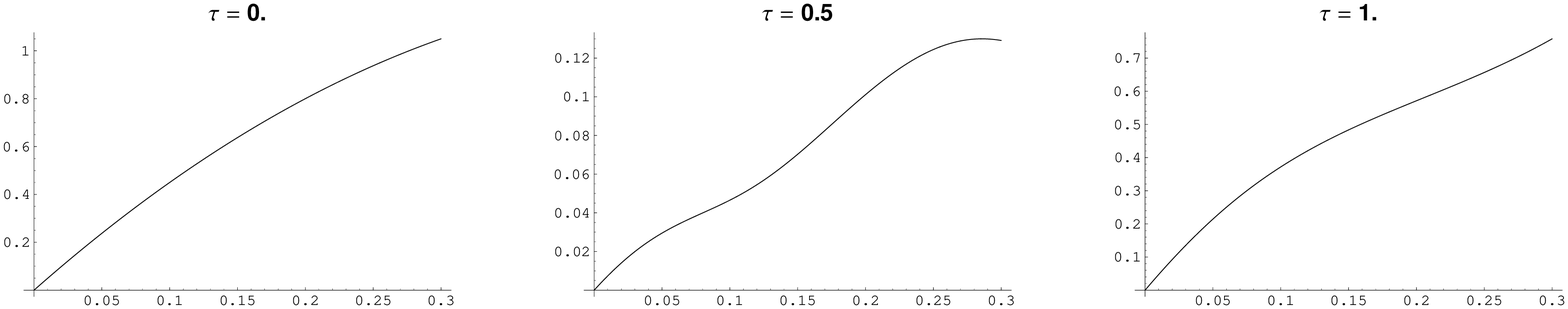}
	\caption{\small{\it Profiles of scalar speed for three value of $\tau$, $0 \leq s \leq 0.3$ and $\mathbb{R}e=10$.}} 
	\end{center}
\end{figure}

\begin{figure}
	\begin{center}
	\includegraphics[width=13cm]{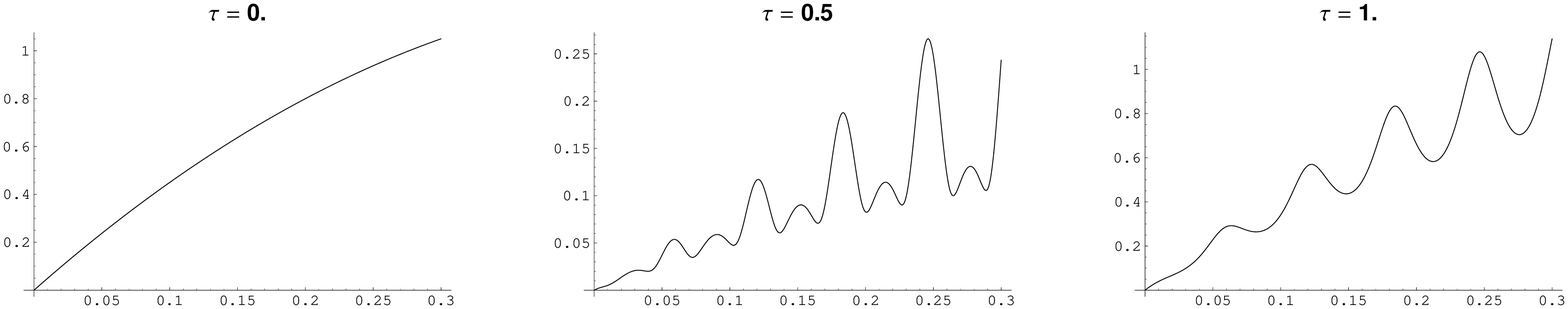}
	\caption{\small{\it Profiles of scalar speed for three value of $\tau$, $0 \leq s \leq 0.3$ and $\mathbb{R}e=100$.}} 
	\end{center}
\end{figure}

\end{document}